\documentclass[twocolumn,superscriptaddress,
showpacs,preprintnumbers,amsmath,amssymb]{revtex4}

\usepackage{graphicx}

\newcommand{\bra}[1]{{\left\langle #1 \right|}}
\newcommand{\ket}[1]{{\left| #1 \right\rangle}}

\newcommand{\tr}{\mbox{$\mathrm{tr}$}}
\newcommand{\C}{\mbox{$\mathcal{C}$}}
\newcommand{\T}{\mbox{$\mathcal{T}$}}
\newcommand{\N}{\mbox{$\mathcal{N}$}}
\newcommand{\I}{\mbox{$\mathcal{I}$}}
\begin{document}

\title{Constraint on teleportation over multipartite pure states}

\author{Jeong San Kim}
\affiliation{
 Institute for Quantum Information Science,
 University of Calgary, Alberta T2N 1N4, Canada
}
\author{Jaewoo Joo}
\affiliation{
 School of Physics and Astronomy,
 University of Leeds, Leeds LS2 9JT, United Kingdom
}
\author{Soojoon Lee}\email{level@khu.ac.kr}
\affiliation{
 Department of Mathematics and Research Institute for Basic Sciences,
 Kyung Hee University, Seoul 130-701, Korea
}

\date{\today}

\begin{abstract}
We first define a quantity
exhibiting the usefulness of bipartite quantum states for teleportation,
called the quantum teleportation capability,
and then investigate its restricted shareability in multi-party quantum systems.
In this work,
we verify that
the quantum teleportation capability has a monogamous property in its shareability
for arbitrary three-qutrit pure states
by employing the monogamy inequality in terms of the negativity.
\end{abstract}

\maketitle

Entanglement is one of the most significant phenomena in
quantum information science as well as quantum mechanics
with no classical counterpart,
and thus has been studied by a lot of scientists for several decades.
Nevertheless, there are still open problems related to entanglement,
especially multipartite entanglement.
In order to understand entanglement more precisely,
we need to explore various kinds of properties about entanglement
and to investigate its useful applications.

Multipartite entanglement has an interesting property, called the monogamy of entanglement,
which means that
quantum states with a specific amount of entanglement cannot be arbitrarily shared
in a multipartite quantum network.
The monogamous property can be considered as
one of the most important features in multipartite entanglement.
In particular, the property can be seen by
the monogamy inequality in terms of the Wootters' concurrence $\C$~\cite{Wootters},
called the Coffman-Kundu-Wootters inequality~\cite{CKW,OV} as follows:
For an arbitrary $n$-qubit pure state $\ket{\psi}_{12\cdots n}$
and its reduced density operators $\rho_{1j}$,
\begin{equation}
\C_{1(2\cdots n)}^2\ge \C_{12}^2+\C_{13}^2+\cdots+\C_{1n}^2,
\label{eq:CKW}
\end{equation}
where $\C_{1(2\cdots n)}=\C(\ket{\psi}_{1(2\cdots n)}\bra{\psi})$
and $\C_{1j}=\C(\rho_{1j})$.
Hence this inequality (\ref{eq:CKW}) shows that
there is an explicit constraint on pure entangled states in multi-qubit systems.

It is clear that
one of the most practical applications of entanglement is teleportation~\cite{BBCJPW},
which is to transmit quantum information between two distant parties
through a classical channel assisted by entanglement.
Thus it is also certain that
the amount of the entanglement to assist the classical channel
determines how reliably teleportation can be performed in the given situation.
On this account, it would be necessary to consider
some quantity to represent the reliable teleportation
for further understanding of entanglement.

Given a two-qudit state $\rho$,
we define the quantity $f$, called the {\it teleportation fidelity}~\cite{Popescu}, as
\begin{equation}
f(\rho)\equiv\int d\xi \bra{\xi}\Lambda_{\rho}(\ket{\xi}\bra{\xi})\ket{\xi},
\label{eq:teleportation_fidelity}
\end{equation}
where the integral is performed
with respect to the uniform distribution $d\xi$ over all one-qudit pure states,
and $\Lambda_{\rho}$ is the standard teleportation scheme over $\rho$
to attain the maximal fidelity.
We remark that $f(\rho)>2/(d+1)$
if and only if $\rho$ is said to be useful for teleportation,
since it has been shown that the classical teleportation
can have a fidelity at most $2/(d+1)$~\cite{Popescu,MP,Horodeckis1}.
Thus, we can define a quantity showing the usefulness for teleportation as
\begin{equation}
\T(\rho)\equiv\max\left\{\frac{(d+1)f(\rho)-2}{d-1},0\right\},
\label{eq:teleportation_capability}
\end{equation}
which we here call the {\em quantum teleportation capability} of a given two-qudit state $\rho$.
Then it is clear that $0\le \T(\rho)\le 1$,
$\T(\rho)=0$ if and only if $\rho$ is said to have no quantum teleportation capability,
and $\T(\rho)=1$ if and only if $\rho$ is said to have full quantum teleportation capability.

One well-known relation between entanglement and teleportation
is that
$\C(\ket{\phi}\bra{\phi})=\T(\ket{\phi}\bra{\phi})$
for all two-qubit pure states $\ket{\phi}$,
and $\C(\rho)\ge\T(\rho)$
for all two-qubit mixed states $\rho$.
It follows from the inequality~(\ref{eq:CKW}) that,
for any $n$-qubit pure state $\ket{\psi}_{12\cdots n}$
and its marginal density matrices $\rho_{1j}$,
\begin{equation}
\T_{1(2\cdots n)}^2\ge \T_{12}^2+\T_{13}^2+\cdots+\T_{1n}^2,
\label{eq:monogamy_ineq}
\end{equation}
where $\T_{1(2\cdots n)}=\T(\ket{\psi}_{1(2\cdots n)}\bra{\psi})$
and $\T_{1j}=\T(\rho_{1j})$.
In other words,
the quantum teleportation capability
also has a monogamous property as a multipartite constraint of teleportation
in the case of $n$-qubit pure states,
whereas the monogamy inequality~(\ref{eq:monogamy_ineq})
does not hold in general for $n$-qubit mixed states~\cite{LP}.
Then one could naturally ask a question as follows:
Does the quantum teleportation capability have such a multipartite constraint,
called the monogamy inequality,
in higher-dimensional quantum systems?

In this work, we present a positive answer to the above question,
and verify that the monogamy inequality in terms of the quantum teleportation capability
holds for numbers of three-qutrit pure states.


Our main idea is very simple.
First, note that
it is generally difficult to calculate
the teleportation fidelity or the quantum teleportation capability of a given two-qudit state,
since the maximal fidelity for all standard teleportation schemes over the same state
should be computed to complete the calculation.
Thus we need another quantity to have some specific relation with the quantum teleportation capability
and to be computable as well.

We here suggest the negativity as such a quantity for the following reason.
For a two-qudit state $\rho_{AB}$,
its (normalized) negativity $\N(\rho_{AB})$~\cite{LKPL,VidalW,LCOK} is defined as
\begin{equation}
\N(\rho_{AB})\equiv \frac{\|\rho_{AB}^{T_B}\|-1}{d-1},
\label{eq:negativity}
\end{equation}
where $\|\cdot\|$ is the trace norm,
and $\rho_{AB}^{T_B}$ is the partial transposition of $\rho_{AB}$.
Remark that
\begin{equation}
\N(\ket{\phi}\bra{\phi})=\T(\ket{\phi}\bra{\phi})
\label{eq:N_T_pure}
\end{equation}
for all two-qudit pure states $\ket{\phi}$
and
\begin{equation}
\N(\rho)\ge\T(\rho)
\label{eq:N_T_mixed}
\end{equation}
for all two-qudit mixed states $\rho$~\cite{VidalW}.
Hence,
in order to prove that
the quantum teleportation capability has a monogamous property
for multipartite pure states,
it suffices to show
that the monogamy inequality in terms of the negativity holds
for pure entangled states in multipartite quantum systems,
that is,
for any $n$-qudit pure state $\ket{\psi}_{12\cdots n}$
and its reduced density matrices $\rho_{1j}$,
\begin{equation}
\N_{1(2\cdots n)}^2\ge \N_{12}^2+\N_{13}^2+\cdots+\N_{1n}^2,
\label{eq:N_monogamy_ineq}
\end{equation}
where $\N_{1(2\cdots n)}=\N(\ket{\psi}_{1(2\cdots n)}\bra{\psi})$
and $\N_{1j}=\N(\rho_{1j})$,
since the inequality~(\ref{eq:N_monogamy_ineq}) directly implies
the monogamy inequality in terms of the quantum teleportation capability.

For simplicity, we consider only three-qutrit pure states in this work.
It has been shown~\cite{Ou,KS} that
there are three-qutrit pure states to violate the monogamy inequality
in terms of the concurrence, such as the states
\begin{equation}
\ket{\mathrm{Ou}}\equiv\frac{1}{\sqrt{6}}
\left(\ket{012}-\ket{021}+\ket{120}-\ket{102}+\ket{201}-\ket{210}\right)
\label{eq:Ou}
\end{equation}
and
\begin{equation}
\ket{\mathrm{KS}}\equiv\frac{1}{\sqrt{6}}
\left(\sqrt{2}\ket{010}+\sqrt{2}\ket{101}+\ket{200}+\ket{211}\right).
\label{eq:KS}
\end{equation}

However, it was also known~\cite{KDS} that
the above two states still satisfy the monogamy inequality
in terms of the convex-roof extended negativity (CREN)~\cite{LCOK}.
Since the CREN value of a given mixed state is not less than its negativity value
by the convexity of the negativity,
we can readily obtain that
the monogamy inequality in terms of the negativity
also holds for
both states $\ket{\mathrm{Ou}}$ in Eq.~(\ref{eq:Ou}) and $\ket{\mathrm{KS}}$ in Eq.~(\ref{eq:KS}).
It follows that the negativity (or the CREN) seems to be an entanglement measure
to show a monogamous property of multipartite entanglement,
while the concurrence does not reveal the monogamy of entanglement
in higher-dimensional quantum systems.

Furthermore, it can be shown that
the monogamy inequality in terms of the negativity still holds
for coherent superpositions of several known three-qutrit pure states~\cite{WO}.
For example, we consider
\begin{eqnarray}
\ket{\mathrm{Ou}_p}_{123}&\equiv&\sqrt{p}\ket{\mathrm{Ou}}_{123}+\sqrt{1-p}\ket{000}_{123},
\label{eq:Ou_S}\\
\ket{\mathrm{KS}_p}_{123}&\equiv&\sqrt{p}\ket{\mathrm{KS}}_{123}+\sqrt{1-p}\ket{222}_{123}.
\label{eq:KS_S}
\end{eqnarray}

{\em Example 1: $\ket{\mathrm{Ou}_p}_{123}$.}
It is straightforward to calculate the reduced density matrices
\begin{eqnarray}
\rho_1^{\mathrm{Ou}}&=&\tr_{23}(\ket{\mathrm{Ou}_p}_{123}\bra{\mathrm{Ou}_p})
\nonumber\\
&=&\frac{p}{3}\I + (1-p)\ket{0}\bra{0},
\label{eq:rho1_Ou}\\
\rho_{ij}^{\mathrm{Ou}}&=&\tr_{k}(\ket{\mathrm{Ou}_p}_{123}\bra{\mathrm{Ou}_p})\nonumber \\
&=&\frac{p}{3}\left(\ket{x}\bra{x}+\ket{y}\bra{y}\right)
+ \left(1-\frac{2p}{3}\right)\ket{z}\bra{z},\nonumber \\
\end{eqnarray}
where $\I$ is the $3\times 3$ identity matrix,
$\{i,j,k\}=\{1,2,3\}$, and
\begin{eqnarray}
\ket{x}&=&\frac{\ket{01}-\ket{10}}{\sqrt{2}},\nonumber\\
\ket{y}&=&\frac{\ket{20}-\ket{02}}{\sqrt{2}},\nonumber\\
\ket{z}&=&\sqrt{\frac{3}{3-2p}}
\left(\frac{\sqrt{p}\left(\ket{12}-\ket{21}\right)}{\sqrt{6}}+\sqrt{1-p}\ket{00}\right).\nonumber\\
\label{eq:xyz}
\end{eqnarray}
By tedious calculations, we have
\begin{eqnarray}
\N_{1(23)}&=&\frac{1}{3}\left(p+2\sqrt{p(3-2p)}\right),\nonumber\\
\N_{1j}&=&
\frac{1}{6}\left(\left|p+\sqrt{6p(1-p)}\right|+\left|p-\sqrt{6p(1-p)}\right|\right)\nonumber\\
&&+\frac{1}{12}\left(\sqrt{33p^2-60p+36}+3p-6\right),
\label{eq:N_123_Ou}
\end{eqnarray}
and hence obtain that
if $0\le p\le 6/7$ then
\begin{widetext}
\begin{eqnarray}
\N_{1(23)}^2-\N_{12}^2-\N_{13}^2
&=&\left(\frac{p+2\sqrt{p(3-2p)}}{3}\right)^2
-2\left(\frac{\sqrt{33p^2-60p+36}+4\sqrt{6p(1-p)}+3p-6}{12}\right)^2,
\label{eq:MoT01_Ou}
\end{eqnarray}
and if $6/7\le p\le 1$ then
\begin{eqnarray}
\N_{1(23)}^2-\N_{12}^2-\N_{13}^2
&=&\left(\frac{p+2\sqrt{p(3-2p)}}{3}\right)^2
-2\left(\frac{\sqrt{33p^2-60p+36}+7p-6}{12}\right)^2.
\label{eq:MoT02_Ou}
\end{eqnarray}
\end{widetext}
It can be directly shown that
both values in Eq.~(\ref{eq:MoT01_Ou}) and Eq.~(\ref{eq:MoT02_Ou}) are nonnegative.
This implies that
$\ket{\mathrm{Ou}_p}_{123}$ satisfies the monogamy inequality in terms of the negativity.

{\em Example 2: $\ket{\mathrm{KS}_p}_{123}$.}
It is also straightforward to calculate the reduced density matrices
\begin{eqnarray}
\rho_1^{\mathrm{KS}}&=&\tr_{23}(\ket{\mathrm{KS}_p}_{123}\bra{\mathrm{KS}_p})\nonumber\\
&=&\frac{p}{3}\I + (1-p)\ket{2}\bra{2},
\label{eq:rho1_KS}\\
\rho_{1j}^{\mathrm{KS}}&=&\tr_{k}(\ket{\mathrm{KS}_p}_{123}\bra{\mathrm{KS}_p})\nonumber \\
&=&\frac{p}{2}\left(\ket{\alpha}\bra{\alpha}+\ket{\beta}\bra{\beta}\right)
+ (1-p)\ket{22}\bra{22},\nonumber \\
\end{eqnarray}
where 
$\{j,k\}=\{2,3\}$, and
\begin{eqnarray}
\ket{\alpha}&=&\frac{\sqrt{2}}{\sqrt{3}}\ket{01}+\frac{1}{\sqrt{3}}\ket{20},\nonumber\\
\ket{\beta}&=&\frac{\sqrt{2}}{\sqrt{3}}\ket{10}+\frac{1}{\sqrt{3}}\ket{21}.
\label{eq:alphabeta}
\end{eqnarray}
Thus we have
\begin{eqnarray}
\N_{1(23)}&=&\frac{1}{3}\left(p+2\sqrt{p(3-2p)}\right),\nonumber\\
\N_{1j}&=&\frac{\sqrt{2}p}{3},
\label{eq:N_123_KS}
\end{eqnarray}
and hence
\begin{eqnarray}
\N_{1(23)}^2-\N_{12}^2-\N_{13}^2
&=&\frac{12p-11p^2+4p\sqrt{p(3-2p)}}{9}\nonumber\\
&\ge& 0,
\label{eq:MoT01_KS}
\end{eqnarray}
that is, $\ket{\mathrm{KS}_p}_{123}$ also satisfies the monogamy inequality in terms of the negativity.

Finally, we numerically verify that the monogamy inequality in terms of the negativity holds
for numbers of three-qutrit pure states.

\begin{figure}
\includegraphics[angle=0,scale=1,width=\linewidth]{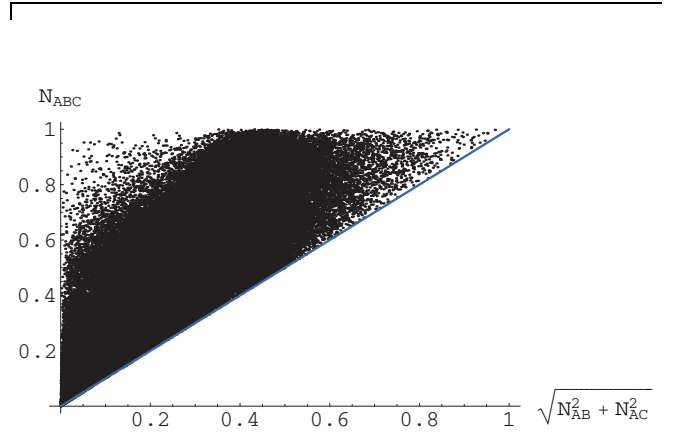}
\caption{\label{Fig:numerical}
Monogamy relation $\N_{A(BC)}^2\ge \N_{AB}^2+\N_{AC}^2$
in terms of the negativity on three-qutrit states:
The blue line represents $\N_{A(BC)}= \sqrt{\N_{AB}^2+\N_{AC}^2}$,
and the black points with coordinates $\left(\sqrt{\N_{AB}^2+\N_{AC}^2},\N_{A(BC)}\right)$
represent possible negativity values
for numbers of three-qutrit pure states.
}
\end{figure}

{\em Example 3: Numerical evidence.}
From the multipartite generalization of the Schmidt decomposition in Ref.~\cite{CHS},
it has been known that
any three-qutrit pure state $\ket{\Psi}_{ABC}$ can be expressed as
\begin{equation}
\ket{\Psi}_{ABC}=\sum_{i,j,k=0}^2 c_{ijk}\ket{i,j,k}_{ABC},
\label{eq:Schmidt3}
\end{equation}
where the coefficients $c_{ijk}$ have the following properties:
\begin{enumerate}
\item $c_{jii} = c_{iji} = c_{iij} = 0$ if $0\le i < j \le 2$;
\item $c_{ijk}$ is real and non-negative if at most one of $i$, $j$, and $k$ differs from $2$;
\item $|c_{iii}| \ge |c_{jkl}|$ if $i \le \min\{j,k,l\}$.
\end{enumerate}
Thus, for numbers of randomly chosen three-qutrit pure states
of the form in Eq.~(\ref{eq:Schmidt3}),
we can numerically compare the values of $\N_{A(BC)}$ and $\sqrt{\N_{AB}^2+\N_{AC}^2}$,
and can carefully conclude that
the inequality $\N_{A(BC)}\ge \sqrt{\N_{AB}^2+\N_{AC}^2}$ holds
for an arbitrary three-qutrit pure state,
as seen in FIG.~\ref{Fig:numerical}.
In other words, a randomly chosen three-qutrit pure state
satisfies the monogamy inequality in terms of the negativity.

The above analytic and numerical results imply that
the monogamy inequality in terms of the negativity
holds for almost all three-qutrit pure states,
and thus
the monogamy inequality in terms of the quantum teleportation capability
is also satisfied for three-qutrit pure states
by exploiting the relation between the negativity and the quantum teleportation capability
in Eq.~(\ref{eq:N_T_pure}) and the inequality~(\ref{eq:N_T_mixed}).
This provides us with a highly probable conjecture that
the quantum teleportation capability has a monogamous property
even for all multipartite qudit pure states
as well as for three-qutrit pure states.

In summary,
we have defined the quantum teleportation capability of a given bipartite state
as a quantity exhibiting the usefulness for teleportation over the state,
and have verified that
the quantum teleportation capability has the monogamous property
for numbers of three-qutrit pure states
by showing the monogamy inequality in terms of the negativity for those states.
Therefore, we have obtained a strong evidence to show that
quantum states with a specific amount of the quantum teleportation capability
cannot be arbitrarily shared in a multipartite quantum network.

This work was supported by the IT R\&D program of the Ministry of Knowledge Economy
[Development of Privacy Enhancing Cryptography on Ubiquitous Computing Environment].
JSK was supported by {\it i}CORE, MITACS and USARO,
JJ acknowledges support by QESSENCE,
and SL was supported by Basic Science Research Program
through the National Research Foundation of Korea (NRF)
funded by the Ministry of Education, Science and Technology (Grant No.~2009-0076578).


\begin{thebibliography}{1}
%
\bibitem{Wootters} W. K.~Wootters,
Phys. Rev. Lett. {\bf 80}, 2245 (1998).
%
\bibitem{CKW}
V.~Coffman, J.~Kundu, and W. K.~Wootters,
Phys. Rev. A {\bf 61}, 052306 (2000).
%
\bibitem{OV}
T. J.~Osborne and F.~Verstraete,
Phys. Rev. Lett. {\bf 96}, 220503 (2006).
%
\bibitem{BBCJPW}
C. H.~Bennett, G.~Brassard, C.~Cr\'{e}peau, R.~Jozsa, A.~Peres, and W. K.~Wootters,
Phys. Rev. Lett. {\bf 70}, 1895 (1993).
%
\bibitem{Popescu}
S. Popescu,
Phys. Rev. Lett. {\bf 72}, 797 (1994).
%
\bibitem{MP}
S.~Massar and S.~Popescu,
Phys. Rev. Lett. {\bf 74}, 1259 (1995).
%
\bibitem{Horodeckis1}
R.~Horodecki, M.~Horodecki, and P.~Horodecki,
Phys. Lett. A {\bf 222}, 21 (1996);
M.~Horodecki, P.~Horodecki, and R.~Horodecki,
Phys. Rev. A {\bf 60}, 1888 (1999).
%
\bibitem{LP}
S.~Lee and J.~Park,
Phys. Rev. A {\bf 79}, 054309 (2009).
%
\bibitem{LKPL}
J.~Lee, M. S.~Kim, Y. J.~Park, and S.~Lee,
J. Mod. Opt. {\bf 47}, 2151 (2000).
%
\bibitem{VidalW}
G.~Vidal and R. F.~Werner,
Phys. Rev. A {\bf 65}, 032314 (2002).
%
\bibitem{LCOK}
S.~Lee, D. P.~Chi, S. D.~Oh and J.~Kim,
Phys. Rev. A {\bf 68}, 062304 (2003).
%
\bibitem{Ou}
Y.-C.~Ou,
Phys. Rev. A {\bf 75}, 034305 (2007).
%
\bibitem{KS}
J. S.~Kim and B. C.~Sanders,
J. Phys. A {\bf 41}, 495301 (2008).
%
\bibitem{KDS}
J. S.~Kim, A.~Das, and B. C.~Sanders,
Phys. Rev. A {\bf 79}, 012329 (2009).
%
\bibitem{WO}
In Refs.~\cite{KS,KDS}, it has already been shown that
any coherent superposition of a generalized W-class state and the state $\ket{000}$
satisfies
the monogamy inequality in terms of the CREN,
and hence
the monogamy inequality in terms of the negativity also holds
for the coherent superpostion
by the convexity of the negativity.
%
\bibitem{CHS}
H. A.~Carteret, A.~Higuchi, and A.~Sudbery,
J. Math. Phys. {\bf 41}, 7932 (2000).
%
\end{thebibliography}
\end{document}